\documentclass[a4paper,11pt]{article}
\usepackage{graphicx,amssymb,bm,latexsym,amsmath}
\newcommand{\bea}{\begin{eqnarray}}
\newcommand{\ena}{\end{eqnarray}}
\newcommand{\vs}[1]{\vspace{#1 mm}}
\newcommand{\hs}[1]{\hspace{#1 mm}}
\renewcommand{\a}{\alpha}
\renewcommand{\b}{\beta}
\renewcommand{\c}{\gamma}
\newcommand{\G}{\Gamma}
\renewcommand{\d}{\delta}
\newcommand{\e}{\epsilon}
\newcommand{\s}{\sigma}
\renewcommand{\t}{\theta}

\newcommand{\la}{\lambda}
\newcommand{\pa}{\partial}

\newcommand{\nn}{\nonumber\\}
\newcommand{\p}[1]{(\ref{#1})}
\newcommand{\lan}{\langle}
\newcommand{\ran}{\rangle}

\newcommand{\tg}{\tilde g}

\begin{document}

\begin{titlepage}

\begin{flushright}
KU-TP 058 \\
%\today
\end{flushright}

\vs{10}
\begin{center}
{\Large\bf Renormalization of Higher Derivative Quantum Gravity
Coupled to a Scalar with Shift Symmetry}
\vs{15}

{\large
Kenji Muneyuki\footnote{e-mail address: kenji.muneyuki@kindai.ac.jp}
and
Nobuyoshi Ohta\footnote{e-mail address: ohtan@phys.kindai.ac.jp}} \\
\vs{10}
{\em Department of Physics, Kinki University,
Higashi-Osaka, Osaka 577-8502, Japan}

\vs{15}
{\bf Abstract}
\end{center}

It has been suggested that higher-derivative gravity theories coupled to a scalar field with
shift symmetry may be an important candidate for a quantum gravity.
We show that this class of gravity theories are renormalizable in $D=3$ and
4 dimensions.

\end{titlepage}
\newpage
\setcounter{page}{2}

\section{Introduction}

It is one of the long standing problems in theoretical physics to
construct quantum theory of gravity.
It has been known for some time that gravity is renormalizable in four dimensions
if one includes higher derivative terms~\cite{Stelle}.
However, the unitarity of the theory, which is one the most important properties of
any physical theory, is not preserved. So the theory has not been taken very seriously.
The compatibility of unitarity and renormalizability has also been studied in \cite{MO}
for three-dimensional theory which could be unitary for judicious choice of parameters.
It turned out that the unitarity and the renormalizability are incompatible.

Recently a very interesting suggestion has been made that the time may be
an emergent notion~\cite{Muk}. The idea starts with four-derivative theory of gravity
coupled to a scalar field with shift symmetry with Euclidean signature.
The quadratic terms of the scalar field also have four derivatives, so that
its scaling dimension is zero. It was assumed that the theory is renormalizable,
but the low-energy effective theory is described by the Einstein theory
together with the four-derivative scalar theory.
It was then shown that this low-energy effective theory is equivalently
described by a Lorentzian action. In this way it was suggested that
the low-energy theory becomes Lorentzian but the theory at the short distance is
described by a Riemannian (locally Euclidean) theory without the notion of
time. If true, this may be a resolution of the ghost problem in the above
renormalizable theory of gravity.

It is necessary to consider such higher derivative terms in gravity since the string theory,
the possible candidate of quantum gravity, predicts that such terms do exist.
Given this fact, we should also take higher derivatives on scalar fields into account
if we further consider scalar fields, and naturally we are led to the class of theories
we consider.

There are several points that have to be confirmed for the above scenario to work.
The obvious and first problem is to explicitly check whether such a theory is really
renormalizable or not.
Though the above quadratic gravity is shown to be renormalizable even in the
presence of a minimal scalar field, i.e. with only the usual kinetic term with second derivative
for the scalar~\cite{Stelle}, it has to be checked if the theory remains renormalizable
with additional higher derivative terms. One may think that it is obvious when
higher derivative kinetic terms are introduced because they improve the convergence
of Feynmann diagrams.
However it is not so because such terms also introduce higher derivative interactions
in the presence of gravity.

More importantly, if the renormalizability is proved, it has to be seen whether
the theory with these higher derivative terms reduces to desirable low-energy effective
theory with the above property. For this purpose, one has to study the renormalization
group and examine the UV and IR fixed points.
For such discussions for higher derivative gravity, see \cite{Av,BS,CP,CPR,LR,BMS,PS,Ohta2}.
This would be next step.

In this paper we take the first step in this direction and examine the renormalization
property of the theories.
There is not much difference between the theories defined for Lorentzian and Euclidean
signature in our perturbative approach. We can simply derive propagators and discuss
power counting and so on, as usual.
We examine whether the theories in $D=3,4$ and 5 dimensions are power counting renormalizable
or not, and show that these theories are super-renormalizable, renormalizable and one-loop
renormalizable, respectively, in these dimensions. Beyond five dimensions,
the theory is not renormalizable.

There are several reasons why we consider not only $D=4$ dimensions but also $D=3$ and 5
dimensions.
First of all, the gravity theories do not have any dynamics in three-dimensional Einstein theory,
but acquire interesting dynamics when higher-derivative terms are added.
They are simple but interesting enough since they can be unitary~\cite{BHT1,Deser,GST,BHT2,Ohta1}.
On the other hand, higher-dimensional gravities are important since string theories live
in higher dimensions and there are several interesting subjects explored in the context of
extra dimensions.
Gravity theories in more than five dimensions cannot be renormalizable in the usual perturbative
approach around the flat Minkowski space even if we add further higher order curvature terms.
So these are the dimensions we are most interested in.

\section{Higher Derivative Gravity}

Let us consider the action of higher derivative gravity coupled to a scalar field $\phi$
with shift symmetry $\phi \to \phi\; +$ constant.
We demand that the theory respect the $Z_2$ symmetry under
$\phi \to -\phi$ as well as the four-dimensional parity $x^\mu \to -x^\mu$,
and contain terms with only up to four derivatives.
The action takes the form~\cite{Muk}
\begin{align}
S =& \int d^D x \sqrt{- g} \Big[
\frac{1}{\kappa^2} \Big( R + \a R^2 +\b R_{\mu\nu}^2 + \c R_{\mu\nu\rho\la}^2 \Big)
+ Z_0 (\nabla_\mu \phi)^2
+ Z_1 R (\nabla_\mu \phi)^2
\nn &
+ Z_2 R^{\mu\nu}\nabla_\mu \phi \nabla_\nu \phi
+ Z_3 (g^{\mu\nu}\nabla_\mu \phi \nabla_\nu \phi)^2
+ Z_4 (\Box \phi)^2
+ Z_5 (\nabla_\mu \nabla_\nu \phi)^2
\Big]
\nn
=& \int d^D x\; \mathcal{L}_{GMG+\phi}
,
\label{action}
\end{align}
where $\kappa^2$ is the $D$-dimensional gravitational constant,
$\a, \b, \c$ and $Z_i$'s ($i=0,\ldots,5$) are constants.
The last term can be set to zero because it can be absorbed into
other terms upon partial integration. Henceforth we set $Z_5=0$.

This is the theory that we examine. Though we should consider the theory in Riemannian
geometry with Euclidean signature, there is not much difference if we discuss Lorentzian
case in our discussions of renormalization.
So in what follows, we discuss this as if the spacetime is the usual Minkowski space.
Also, we have written down the above action only for a single scalar field for simplicity,
but it is straightforward to extend the theory with several scalar fields.

\subsection{Propagator}
\label{prop}

We define the fluctuation around the Minkowski background by
\bea
\tilde g^{\mu\nu}\equiv \sqrt{-g}\, g^{\mu\nu}
= \eta^{\mu\nu} + \kappa h^{\mu\nu}.
\label{fluc}
\ena
For simplicity, we set $\kappa=1$.
Substituting \p{fluc} into our action~\p{action}, we find the quadratic
term is given by
\bea
{\cal L}_2 \hs{-2}&=&\hs{-2} \frac{1}{4} h^{\mu\nu} \Big[ \{(\b+4\c)\Box+1\} P^{(2)}
+ \frac{\{(D-1)(4\a+\b)+\b+4\c\}\Box-(D-2)}{(D-2)^2} \{P^{(0,s)} \nn
&& \hs{5}
+ (D-1) P^{(0,w)} +\sqrt{D-1}(P^{(0,sw)}+P^{(0,ws)}) \} \Big]_{\mu\nu,\rho\s}\Box h^{\rho\s}
+ \phi (Z_4 \Box+Z_0)\Box \phi,
\ena
where we have defined the projection operators as
\bea
P^{(2)}_{\mu\nu,\rho\s} \hs{-2}&=&\hs{-2} \frac12 \Big(\t_{\mu\rho}\t_{\nu\s}
+\t_{\mu\s}\t_{\nu\rho} -\frac{2}{D-1}\t_{\mu\nu}\t_{\rho\s}\Big), \nn
P^{(1)}_{\mu\nu,\rho\s} \hs{-2}&=&\hs{-2} \frac12(\t_{\mu\rho}\omega_{\nu\s}
+\t_{\mu\s}\omega_{\nu\rho}+\t_{\nu\rho}\omega_{\mu\s} + \t_{\nu\s}\omega_{\mu\rho}), \nn
P^{(0,s)}_{\mu\nu,\rho\s} \hs{-2}&=&\hs{-2} \frac{1}{D-1} \t_{\mu\nu}\t_{\rho\s}, ~~
P^{(0,w)}_{\mu\nu,\rho\s} =  \omega_{\mu\nu}\omega_{\rho\s},\nn
P^{(0,sw)}_{\mu\nu,\rho\s} \hs{-2}&=&\hs{-2} \frac1{\sqrt{D-1}} \t_{\mu\nu}\omega_{\rho\s},~~
P^{(0,ws)}_{\mu\nu,\rho\s} = \frac1{\sqrt{D-1}} \omega_{\mu\nu}\t_{\rho\s},
\ena
with
\bea
\t_{\mu\nu} = \eta_{\mu\nu}-\frac{\pa_\mu \pa_\nu}{\Box}, \qquad
\omega_{\mu\nu} = \frac{\pa_\mu \pa_\nu}{\Box}.
\ena
$P^{(2)}, P^{(1)}, P^{(0,s)}$ and $P^{(0,w)}$ are the projection operators onto spin
2, 1 and 0 parts, and they satisfy the completeness relation
\bea
( P^{(2)}+P^{(1)}+ P^{(0,s)}+ P^{(0,w)})_{\mu\nu,\rho\s}
= \frac12 (\eta_{\mu\rho} \eta_{\nu\s}+ \eta_{\mu\s} \eta_{\nu\s}),
\label{complete}
\ena
on the symmetric second-rank tensors.
Note that $\c$ can be eliminated by the shift $\a\to\a-\c$ and $\b\to\b-4\c$.
(We see that this is true for any dimension in this order, but is valid only for
$D=3$ and 4 at the nonlinear level.)
However we keep $\c$ here since $\c$ is expected to be relevant in dimensions higher than 4.

The BRST transformation for the fields is found to be
\bea
\d_B g_{\mu\nu} &=& -\d \la [ g_{\rho\nu}\pa_\mu c^\rho + g_{\rho\mu}\pa_\nu c^\rho
 + \pa_\rho g_{\mu\nu} c^\rho] , \nn
\d_B c^\mu &=& -\d\la c^\rho \pa_\rho c^\mu, \nn
\d_B \bar c_\mu &=& i \d\la\, B_\mu,\nn
\d_B B_\mu &=& 0, \nn
\d_B \phi &=& -\d\la c^\rho \pa_\rho \phi,
\label{brst}
\ena
which is nilpotent. Here $\d\la$ is an anticommuting parameter.
We use the same gauge fixing as \cite{Stelle} using $\tg^{\mu\nu}$,
whose BRST transformation is given by
\bea
\d_B \tg^{\mu\nu} = \d\la (
\tg^{\mu\rho}\pa_\rho c^\nu + \tg^{\nu\rho}\pa_\rho c^\mu
-\tg^{\mu\nu}\pa_\rho c^\rho - \pa_\rho\tg^{\mu\nu} c^\rho)
\equiv \d\la {\cal D}^{\mu\nu}{}_\rho c^\rho.
\ena
The gauge fixing term and Faddeev-Popov (FP) ghost terms are concisely written as
\bea
{\cal L}_{GF+FP} &=& i \d_B [\bar c_\mu (\pa_\nu h^{\mu\nu}-\frac{a}{2}B^\mu)]/\d\la \nn
&=& - B_\mu \pa_\nu h^{\mu\nu}
- i \bar c_\mu \pa_\nu {\cal D}^{\mu\nu}{}_\rho c^\rho +\frac{a}{2} B_\mu B^\mu,
\label{gfgh}
\ena
where $a$ is a gauge parameter and the indices are raised and lowered with
the flat metric.

The simplest way to read off the propagator for the graviton is to first eliminate
the auxiliary field $B^\mu$ and look at the quadratic part. We find that it is given by
\bea
{\cal L}_{2,t} \hs{-2}&=&\hs{-2} \frac{1}{4} h^{\mu\nu} \Big[ \{(\b+4\c)\Box+1\} P^{(2)}
+\frac{1}{a}P^{(1)}
+ \frac{\{4(D-1)\a+D\b+4\c\}\Box-(D-2)}{(D-2)^2} \left\{P^{(0,s)} \right. \nn
&& \hs{10}
\left. + (D-1) P^{(0,w)} +\sqrt{D-1}\left(P^{(0,sw)}+P^{(0,ws)}\right) \right\}
+\frac{2}{a}P^{(0,w)} \Big]_{\mu\nu,\rho\s}\Box h^{\rho\s}
\label{quad}
\ena
Using the completeness property~\p{complete} and the orthogonality of the projection operators,
we find the propagators are given by
\bea
D_{\mu\nu,\rho\s}^h (k) \hs{-2}&=&\hs{-2} \frac{4}{(2\pi)^D}
\Big[ \frac{P^{(2)}}{k^2\{(\b+4\c) k^2-1\}}
+ \frac{(D-2)^2 P^{(0,s)}}{ k^2 [ \{4(D-1)\a+D\b+4\c\}k^2+ D-2 ]} \nn
&& \hs{-5}-\frac{a}{2 k^2} \left\{2P^{(1)} +(D-1) P^{(0,s)} + P^{(0,w)}
-\sqrt{D-1}\left(P^{(0,sw)}+P^{(0,ws)}\right) \right\} \Big]_{\mu\nu,\rho\s}, \\
D^\phi (k) \hs{-2}&=&\hs{-2} \frac{1}{(2\pi)^D} \frac{1}{k^2(Z_4 k^2 + Z_0)}.
\label{propagator}
\ena
We shall take the Landau gauge
given by $a=0$. Since the theory is invariant under the general coordinate transformation,
this does not cause any problem, but simplifies the discussions considerably~\cite{Stelle,MO}.
In this gauge, we have
\bea
\pa_\mu h^{\mu\nu}=0.
\label{landau}
\ena
Note that the above propagator satisfies $k^\mu D_{\mu\nu,\rho\s}^h (k)=0$ in this gauge.
Also, these propagators damp as $k^{-4}$ for large momentum.
The ghost propagator damps as $k^{-2}$, but there is special property in this case.
As a result, as we argue later, the theory becomes renormalizable.

\subsection{Slavnov-Taylor identity}
\label{stid}

If we introduce the Grassmann-odd source $K_{\mu\nu}$ and M, and the Grassmann-even source $L_\mu$,
we have the BRST-invariant action
\bea
&& \hs{-20}
I_{sym}[h_{\mu\nu}, \bar c_\a, c^\b, K_{\mu\nu}, L_\rho, M] \nn
\hs{-2} &=& \hs{-2} \int d^D x[ {\cal L}_{GMG+\phi}+{\cal L}_{GF+FP}
+ K_{\mu\nu}{\cal D}^{\mu\nu}{}_\rho c^\rho - L_\mu c^\nu \pa_\nu c^\mu
- Mc^{\rho}\pa_{\rho}\phi] \nn
& \equiv &\hs{-2}  \int d^D x \; {\cal L}_{sym}.
\ena
The BRST invariance follows from \p{brst}, \p{gfgh} and the nilpotency of
the BRST transformation.

The generating functional of Green's functions is given by
\bea
&& \hs{-20}
Z[J_{\mu\nu}, \bar \eta_\a, \eta^\b, N, K_{\mu\nu}, L_\rho, M] \nn
\hs{-2} &=& \hs{-2} \int [dh][d\phi][d\bar c][dc] \exp\left(i\int d^D x [ {\cal L}_{sym}
+J_{\mu\nu} h^{\mu\nu} +N\phi + \bar\eta_\a c^\a + \bar c_\a \eta^\a] \right) \nn
\hs{-2} &\equiv& \hs{-2} \exp\left( i W[J_{\mu\nu}, \bar \eta_\a, \eta^\b, K_{\mu\nu}, L_\rho, M]
 \right),
\label{generating}
\ena
where $J_{\mu\nu}$ and $N$ ($\bar\eta_\a$ and $\eta^\a$) are Grassmann-even
(Grassmann-odd) sources, respectively.
The BRST invariance of the functional~\p{generating}
\bea
0 = \int [dh][d\phi][d\bar c][dc] \d_B \exp\left(i\int d^D x [ {\cal L}_{sym}
+J_{\mu\nu} h^{\mu\nu} +N\phi + \bar\eta_\a c^\a + \bar c_\a \eta^\a] \right),
\ena
implies that
\bea
\left\lan \int d^D x \left[ J_{\mu\nu} {\cal D}^{\mu\nu}{}_\rho c^\rho
- N c^{\rho} \pa_{\rho} \phi
+ \bar \eta_\mu c^\nu \pa_\nu c^\mu +i\frac{1}{a} \eta_\mu \pa_\nu h^{\mu\nu}
 \right] \right\ran = 0,
\ena
where the field $B_\mu$ is eliminated by its field equation.
This yields the Slavnov-Taylor identity
\bea
\int d^D x \left[ J_{\mu\nu} \frac{\d W}{\d K_{\mu\nu}}
+ N \frac{\d W}{\d M}
 -\bar\eta_\mu \frac{\d W}{\d L_\mu}
+\frac{i}{a} \eta^\mu \pa_\nu \frac{\d W}{\d J_{\mu\nu}}
 \right]=0.
\ena
The equations of motion for the FP ghost is
\bea
\pa_\nu \frac{\d W}{\d K_{\mu\nu}} + i \eta_\mu=0.
\ena
As usual, the effective action is defined by
\bea
&& \hs{-20}
\tilde \G[h^{\mu\nu}, \phi, \bar c_\a, c^\b, K_{\mu\nu}, L_\rho, M] \nn
&\equiv& W[J_{\mu\nu}, \bar \eta_\a, \eta^\b, K_{\mu\nu}, L_\rho, M]
-\int d^D x \left[J_{\mu\nu} h^{\mu\nu} +N\phi +\bar\eta_\a c^\a+\bar c_\a \eta^\a \right].
\label{effective}
\ena
It follows from \p{generating} that
\bea
h^{\mu\nu} = \frac{\d W}{\d J_{\mu\nu}},~~~
\phi = \frac{\d W}{\d N},~~~
c^\mu = \frac{\d W}{\d \bar\eta_\mu}, ~~
\bar c_\mu = - \frac{\d W}{\d \eta^\mu}.
\ena
The relations dual to these are
\bea
J_{\mu\nu} = - \frac{\d \tilde\G}{\d h^{\mu\nu}}, ~~~
N = -\frac{\d \tilde\G}{\d \phi}, ~~~
\bar \eta_\a = \frac{\d \tilde\G}{\d c^\a}, ~~
\eta^\a = -\frac{\d \tilde\G}{\d \bar c_\a}.
\ena
We further define
\bea
\G=\tilde\G+ \int d^D x \frac{1}{2a} (\pa_\nu h^{\mu\nu})^2.
\ena
With the help of the relations
\bea
\frac{\d \G}{\d K_{\mu\nu}}= \frac{\d W}{\d K_{\mu\nu}},~~~
\frac{\d \G}{\d M} = \frac{\d W}{\d M},~~~
\frac{\d \G}{\d L_\mu}= \frac{\d W}{\d L_\mu},
\ena
and the ghost field equation
\bea
\pa^\nu \frac{\d \G}{\d K_{\mu\nu}} - i \frac{\d \G}{\d \bar c^\mu} = 0,
\ena
the Slavnov-Taylor identity reduces to
\bea
\label{stid1}
\int d^D x \left[ \frac{\d \G}{\d h^{\mu\nu}} \frac{\d \G}{\d K_{\mu\nu}}
+ \frac{\d \G}{\d c^\mu} \frac{\d \G}{\d L_\mu}
+ \frac{\d \G}{\d \phi}\frac{\d \G}{\d M}\right]=0,
\ena

The $n$-loop part of the effective action is denoted by $\G^{(n)}$.
The effective action is a sum of these terms:
\bea
\G = \sum_{n=0}^\infty \G^{(n)}.
\ena
Suppose that we have successfully renormalized the effective action up to
$(n-1)$-loop order. Write
\bea
\G^{(n)} = \G^{(n)}_{\rm finite} + \G^{(n)}_{\rm div}.
\ena
If we insert this breakup into Eq.~\p{stid1} and keep only the terms which are
of $n$-loop order, we get
\bea
\int d^D x \left[ \frac{\d \G^{(n)}_{\rm div}}{\d h^{\mu\nu}}\frac{\d \G^{(0)}}{\d K_{\mu\nu}}
+ \frac{\d \G^{(0)}}{\d h^{\mu\nu}}\frac{\d \G^{(n)}_{\rm div}}{\d K_{\mu\nu}}
+ \frac{\d \G^{(n)}_{\rm div}}{\d c^\mu}\frac{\d \G^{(0)}}{\d L_\mu}
+ \frac{\d \G^{(0)}}{\d c^\mu}\frac{\d \G^{(n)}_{\rm div}}{\d L_\mu}
+ \frac{\d \G^{(n)}_{\rm div}}{\d \phi}\frac{\d \G^{(0)}}{\d M}
+ \frac{\d \G^{(0)}}{\d \phi}\frac{\d \G^{(n)}_{\rm div}}{\d M}\right] \nn
= - \int d^D x  \sum_{i=0}^n \left[
\frac{\d \G^{(n-i)}_{\rm finite}}{\d h^{\mu\nu}}\frac{\d \G^{(i)}_{\rm finite}}{\d K_{\mu\nu}}
+ \frac{\d \G^{(n-i)}_{\rm finite}}{\d c^\rho}\frac{\d \G^{(i)}_{\rm finite}}{\d L_\rho}
+ \frac{\d \G^{(n-i)}_{\rm finite}}{\d \phi}\frac{\d \G^{(i)}_{\rm finite}}{\d M}
\right].
\label{stid2}
\ena
Since each term on the right-hand side of \p{stid2} remains finite as $\e\to 0$
in the dimensional regularization, while each term on the left-hand side contains
a factor with a pole in $\e$, each side of the equation must vanish separately.
This leads to
\bea
\int d^D x \left[ \frac{\d \G^{(0)}}{\d K_{\mu\nu}} \frac{\d }{\d h^{\mu\nu}}
+ \frac{\d \G^{(0)}}{\d h^{\mu\nu}}\frac{\d}{\d K_{\mu\nu}}
+ \frac{\d \G^{(0)}}{\d L_\la}\frac{\d }{\d c^\la}
+ \frac{\d \G^{(0)}}{\d c^\la}\frac{\d}{\d L_\la}
+ \frac{\d \G^{(0)}}{\d \phi}\frac{\d }{\d M}
+ \frac{\d \G^{(0)}}{\d M}\frac{\d }{\d \phi}\right] \G^{(n)}_{\rm div}
=0.
\label{stid3}
\ena
This identity will be used in later discussions of renormalizability.

\subsection{Renormalizability}

Under the expansion~\p{fluc}, the Einstein term gives graviton vertices with
two derivatives, and curvature square terms give those with four derivatives.
Compared with the theory without scalar~\cite{MO}, we also have scalar vertices
with four derivatives as well as scalar-graviton vertices with two and four derivatives.

Consider arbitrary Feynmann diagrams. We use the following notations.
\\

$V_{h,2}$: the number of graviton vertices with two derivatives from the $R$ term.

$V_{h,4}$: the number of graviton vertices with four derivatives from the $R^2$ term.

$V_{s,4}$: the number of scalar vertices with four derivatives.

$V_{hs,2}$: the number of graviton-scalar vertices with two derivatives.

$V_{hs,4}$: the number of graviton-scalar vertices with four derivatives.

$V_c$: the number of ghost-antighost-graviton vertices with two derivatives.

$V_K$: the number of $K$-graviton-ghost vertices.

$V_L$: the number of $L$-ghost-ghost vertices.

$V_M$: the number of $M$-graviton-ghost vertices.

$I_h$: the number of internal-graviton propagators.

$I_s$: the number of internal-scalar propagators.

$I_c$: the number of internal-ghost propagators.

$E_h$: the number of external gravitons.

$E_c$: the number of external ghosts.
\\

Since the graviton and scalar propagators behaves as $k^{-4}$ and the FP ghost
propagator as $k^{-2}$,
we are led by the standard power counting to the degree of divergence of
an arbitrary diagram:
\bea
D_{div} &=& D L-4 I_h -4I_s-2 I_c+4 V_{h,4}+2 V_{h,2}+4 V_{s,4}+4V_{hs,4}+2V_{hs,2} \nn
&& +2 V_c+V_K+V_L +V_M.
\ena
Using the relation
\bea
L=I_h+I_c+I_s-(V_{h,4}+V_{h,2}+ V_{s,4}+V_{hs,4}+V_{hs,2}+V_c+V_K+V_L+V_M-1),
\ena
we get
\bea
D_{div} \hs{-2}&=&\hs{-2} D +(D-4)(I_h+I_s-V_{h,4}-V_{s,4}-V_{hs,4})
+(D-2)(I_c-V_{h,2}-V_{hs,2}-V_c) \nn
&& -(D-1)(V_K+V_L+V_M).
\ena
We further use the topological relation
\bea
2V_c+V_K+2V_L+V_M=2I_c+E_c+E_{\bar c},
\ena
to obtain
\bea
D_{div} \hs{-2}&=&\hs{-2}D- (4-D)(I_h+I_s-V_{h,4}-V_{s,4}-V_{hs,4})-(D-2)(V_{h,2}+V_{hs,2})\nn
&& -\; \frac{D}{2} (V_K+V_M) - V_L- \frac{D-2}{2} (E_c+ E_{\bar c}).
\label{pc}
\ena
Now the ghost vertex contained in the FP ghost term in \p{gfgh}, upon partial integration,
can be rewritten as
\bea
i[\pa_\rho \pa_\mu \bar c_\nu \cdot c^\nu h^{\mu\rho}
+\pa_\mu \bar c_\nu \cdot c^\nu \pa_\rho h^{\mu\rho}
+\pa_\mu \bar c_\nu \cdot c^\mu \pa_\rho h^{\nu\rho}].
\ena
In the Landau gauge in which we have \p{landau}, the last two terms do not couple to the
propagator. Also integration by parts in the remaining term can be used to move
the derivative onto the ghost using the gauge condition:
\bea
i \pa_\rho \pa_\mu \bar c_\nu \cdot c^\nu h^{\mu\rho}
\approx i \bar c_\nu \pa_\rho \pa_\mu c^\nu h^{\mu\rho}.
\ena
As a result, in one-particle irreducible (1PI) diagrams, each external ghost and antighost
carries two factors of external momentum~\cite{Stelle,MO}.
The resulting degree of divergence of an arbitrary 1PI diagram is then
\bea
D^{(1PI)}_{div} \hs{-2}&=&\hs{-2} D-(4-D)(I_h+I_s-V_{h,4}-V_{s,4}-V_{hs,4})
-(D-2)(V_{h,2}+V_{hs,2})\nn
&& -\; \frac{D}{2} (V_K+V_M) - V_L- \frac{D+2}{2} (E_c+ E_{\bar c}).
\ena
We note that $I_h+I_s-V_{h,4}-V_{s,4}-V_{hs,4} \geq 0$ for 1PI diagrams,
so most of the contributions are negative for $D\leq 4$.

First, let us concentrate on $D=3$.
The resulting degree of divergence of an arbitrary 1PI diagram is then
\bea
D^{(1PI)}_{div} &=&\hs{-2} 3 - (I_h+I_s-V_{h,4}-V_{s,4}-V_{hs,4}) \nn
&&- (V_{h,2}+V_{hs,2})-\frac32 (V_K+V_M)- V_L- \frac52 (E_c + E_{\bar c}).
\ena
We find that the possible
divergences are restricted; those with external ghosts and antighosts have
$D^{\rm (1PI)}\leq -2$, those with the external $K$ and ghost
$D^{\rm (1PI)} \leq -1$, those with $L$ and two ghosts have $D^{\rm (1PI)} \leq -3$,
and those with external $M$ and ghost have $D^{\rm (1PI)} \leq -1$.
Hence, we have
\bea
\frac{\d \G^{(n)}_{\rm div}}{\d c^\la}
= \frac{\d \G^{(n)}_{\rm div}}{\d K^{\mu\nu}}
= \frac{\d \G^{(n)}_{\rm div}}{\d L_\la}
= \frac{\d \G^{(n)}_{\rm div}}{\d M}=0.
\label{nodiv}
\ena
The Slavnov-Taylor identity~\p{stid3} then reduces to
\bea
\int d^3 x \bigg[\frac{\d \G^{(0)}}{\d K^{\mu\nu}} \frac{\d }{\d \tg^{\mu\nu}}
+\frac{\d \G^{(0)}}{\d M} \frac{\d }{\d \phi
}\bigg]\G^{(n)}_{\rm div}
=0.
\ena
Together with \p{nodiv}, this implies that $\G^{(n)}_{\rm div}$ is gauge invariant.
Consequently $\G^{(n)}_{\rm div}$ are local gauge-invariant functionals of
$\tg^{\mu\nu}$ and $\phi$ with zero and two derivatives (up to three).
This allows only the counterterms of the Einstein and cosmological,
and $(\nabla_{\mu}\phi)^2$ terms. Terms like $\phi^2$  are not allowed due to
the shift symmetry.
Clearly we have divergence at the lower-loop levels. The convergence
property improves as more vertices and internal lines are added,
{\it i.e.} when we go to higher-loop diagrams.
Thus there are only finite numbers of divergent diagrams.
Hence the theory is super-renormalizable.

Next, let us consider $D=4$.
The degree of divergence of an arbitrary 1PI diagram is
\bea
D^{(1PI)}_{div} = 4 - 2(V_{h,2}+V_{hs,2})- 2 (V_K+V_M)- V_L- 3 (E_c + E_{\bar c}).
\ena
We find that the possible divergences are again restricted;
those with external ghosts and antighosts have
$D^{\rm (1PI)}\leq -2$, those with the external $K$ and ghost
$D^{\rm (1PI)} \leq -1$, those with $L$ and two ghosts have $D^{\rm (1PI)} \leq -3$,
and those with external $M$ and ghost have $D^{\rm (1PI)} \leq -1$.
Hence, we have
\bea
\frac{\d \G^{(n)}_{\rm div}}{\d c^\la}
= \frac{\d \G^{(n)}_{\rm div}}{\d K^{\mu\nu}}
= \frac{\d \G^{(n)}_{\rm div}}{\d L_\la}
= \frac{\d \G^{(n)}_{\rm div}}{\d M}=0.
\label{nodiv2}
\ena
The Slavnov-Taylor identity~\p{stid3} then reduces to
\bea
\int d^4 x \bigg[\frac{\d \G^{(0)}}{\d K^{\mu\nu}} \frac{\d }{\d \tg^{\mu\nu}}
+\frac{\d \G^{(0)}}{\d M} \frac{\d }{\d \phi
}\bigg]\G^{(n)}_{\rm div}
=0.
\ena
Together with \p{nodiv2}, this implies that $\G^{(n)}_{\rm div}$ is gauge invariant.
Therefore $\G^{(n)}_{\rm div}$ are local gauge-invariant functionals of $\tg^{\mu\nu}$ and
$\phi$ with zero, two and four derivatives.
This allows only the counterterms that are the same as \p{action} (and cosmological constant).
Thus the theory is renormalizable.
In this case, we find that the divergence is not necessarily restricted to lower loops
since the number of vertices from higher-derivative terms and internal lines do not
affect the degree of divergence and hence there may be divergent diagrams involving
these in higher loops. In this sense, the theory is only renormalizable.

Finally we turn to $D=5$.
The degree of divergence of an arbitrary 1PI diagram is then
\bea
D^{(1PI)}_{div} &=& 5+(I_h+I_s-V_{h,4}-V_{s,4}-V_{hs,4})-3(V_{h,2}+V_{hs,2}) \nn
&& -\frac52 (V_K+V_M)- V_L- \frac72 (E_c + E_{\bar c}).
\ena
Diagrams are more divergent when we go to higher loops since
$I_h+I_s-V_{h,4}-V_{s,4}-V_{hs,4} > 0$, and
the theory in $D=5$ is not renormalizable.
However, for the one-loop diagrams, the second term vanishes and we have only
the same divergences as for $D=4$.
The theory is renormalizable at this level.

Beyond $D=5$, we get more divergences, and the theory is not renormalizable.
It is clear that adding higher curvature terms does not help to improve renormalizability,
because these do not improve the behavior of propagators around Minkowski space
but introduce higher derivative vertices.

\section{Conclusions}

In this paper we have studied whether higher derivative gravities coupled to a scalar field
with shift symmetry in $D=3,4,5$ dimensions are renormalizable or not.
 We have shown that the general theory is (super-)renormalizable in $D=3~\text{and}~4$,
and is not renormalizable in $D=5$.
Theory in $D=5$ is renormalizable in the one-loop calculations,
because $I_h+I_s-V_{h,4}-V_{s,4}-V_{hs,4} = 0$ for one-loop 1PI diagrams.
We have noted that theories in further higher dimensions are not renormalizable
around Minkowski space even if more higher curvature terms are added.
So this analysis exhausts interesting cases in the perturbative approach.

Thus the first step described in the introduction is cleared.
The next issue to be studied is the renormalization group properties of these theories.
In the four-dimensional case, we expect that the coefficients of the terms in the action
have Gaussian fixed point as well as other nontrivial fixed point.
The existence of the fixed points is named asymptotic safety, and
the Gaussian fixed point corresponds to asymptotic freedom.
This has been checked for other type of higher derivative theories~\cite{Av} -- \cite{BMS},
but it should be confirmed in our theory explicitly.
On the other hand, in the low energy, we expect that the higher derivative terms
become irrelevant and Einstein term has a finite fixed point.
It would be interesting to explicitly check if the theory reduces to the Einstein theory
together with the four-derivative scalar theory which is equivalently
described by a Lorentzian action.
We hope to report on this problem elsewhere.

\section*{Acknowledgement}

We would like to thank Shinji Mukohyama
for very useful discussions.
This work was supported in part by the Grant-in-Aid for
Scientific Research Fund of the JSPS (C) No. 24540290
and (A) No. 22244030.

\end{document}